\documentclass[aps,showpacs]{revtex4}

\usepackage{epsfig}
\usepackage{epsf}
\usepackage{bm}                 % for bold math symbols
\usepackage{amsmath}
\usepackage{amssymb}
\usepackage{dcolumn}
\usepackage{graphicx}
\usepackage{xcolor}
\usepackage[small,bf, format=plain, width=.75\textwidth, justification=RaggedRight]{caption}[2008/04/01]
\input amssym.def
\input amssym
\font\script=eusm10
\font\bbold=msbm10
\newcommand{\SH}{\mbox{\script H}}

\newcommand{\BC}{\mbox{\bbold C}}

\newcommand{\BigFig}[1]{\parbox{14pt}{\Huge #1}}
\newcommand{\BigParRight}{\BigFig{]}}
\newcommand{\BigParLeft}{\BigFig{[}}

\begin{document}

\title{Modeling lossy propagation of non-classical light}

\author{Ulvi Yurtsever} \email{ulvi@phys.lsu.edu}
\affiliation{MathSense Analytics, 1273 Sunny Oaks Circle, Altadena, CA 91001 and
\\ Hearne Institute for Theoretical Physics, Louisiana State University,
Baton Rouge, LA 70803}
%\author{Gabriel Durkin} \email{}
%\affiliation{NASA AMES Reserch Center}

\date{\today}

\begin{abstract}

The lossy propagation law (generalization of Lambert-Beer's law for classical radiation
loss) for non-classical, dual-mode entangled states is derived from first principles,
using an infinite-series of beam splitters to model continuous photon loss.
This model is general enough to
accommodate stray-photon noise along the propagation, as well as amplitude attenuation.
An explicit analytical
expression for the density matrix as a function of propagation distance is obtained
for completely general input states with bounded photon number in each mode.
The result is analyzed numerically for various examples of input states. For N00N state
input, the loss of coherence and entanglement
is super exponential as predicted by a number of previous studies.
However, for generic input states, where the coefficients are generated randomly, the decay of
coherence is very different; in fact no worse than the classical Beer-Lambert law. More surprisingly,
there is a plateau at a mid-range interval in propagation distance where the loss is in fact
sub-classical, following which it resumes the classical rate. The qualitative behavior
of the decay of entanglement for two-mode propagation is also analyzed numerically
for ensembles of random states using
the behavior of negativity as a function of propagation distance.

\end{abstract}

\pacs{03.65.Ta, 06.20.Dk, 42.50.Lc, 42.50.St}

\maketitle

We will derive the quantum analogue of the classical Beer-Lambert law for lossy
propagation~\cite{beer law} from first principles by modeling the propagation medium
as a continuum series of linear optical scattering elements. 
The results we report here suggest that the super-exponential propagation loss behavior of N00N
states is highly special, and not likely to be shared by generic entangled states in the larger
Hilbert space of the dual photon channel. The intuition that entanglement
embedded in a general state would decay similarly to the decay
of entanglement in the lossy propagation of a N00N state appears to be faulty.
Assuming even part of the coherent entanglement
that survives during the propagation of a generic
entangled state
can be utilized to produce super-classical phase sensitivity using an appropriate detection
scheme~\cite{quantum interf},
there appear to be many candidate states which are both robust against decoherence and
non-classical enough to achieve significant advantage over classical light
in an optimized quantum sensor architecture.
This is consistent with
and a generalization of previous results involving m-and-m states~\cite{m-m}.

The basic model is illustrated in Figure 1. Here we first analyze the
propagation of single-mode quantum light. The model assumes a series of $M$ identical
beam splitters into which that the input modes are
$a_0 , \, d_1 , \, d_2 , \, \ldots , d_M $, where $a_0$ is the incoming photon mode,
and $d_1 , \, d_2 , \, \ldots , d_M $ are auxiliary channels possibly populated by stray light
or thermal photons, but for the discussion here we will
assume these channels have vacuum inputs. Similarly, the output
modes are $a_M , \, s_1 , \, s_2 , \, \ldots , s_M $, where $a_M$ is the output photon mode
corresponding to the input channel $a_0$,
and $s_1 , \, s_2 , \, \ldots , s_M $ are auxiliary channels modeling scattered
and absorbed light along the propagation medium. The idea is to let the number
of beam splitters $M$ approach infinity in a
controlled manner at the end of the calculation to extract the loss behavior of the
quantum state propagating non-unitarily from the $a_0$ channel to the $a_M$ channel.

\begin{figure}[htp]
%\vspace{0.5in}
\hspace{3.0in}
\centerline{
\input epsf
\setlength{\epsfxsize}{4.1in}%4.200
\setlength{\epsfysize}{3.2in}%3.245
\epsffile{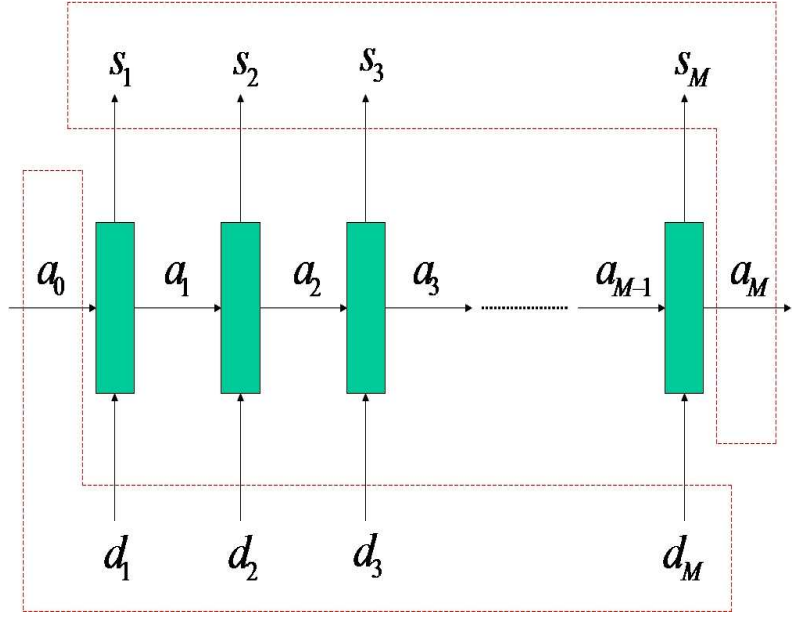}
}
%\vspace{-0.2in}
\caption[figure]{\label{fig:figure1}
Diagram illustrating the series-of-beam-splitters model for propagation loss.}
\vspace{0.10in}
\end{figure}

At each beam splitter $k$ in Fig.\ 1, unitarity requires the relations
\begin{eqnarray}
a_k & = & T \, a_{k-1} + L \, d_k \;,
\nonumber \\
s_k & = & L \, a_{k-1} + T \, d_k \; , \; \; \; k=1,\; 2, \; \ldots , \; M \; ,
\end{eqnarray}
where $T$ and $L$ are the complex transmission and reflection coefficients
and satisfy the unitarity conditions
\begin{eqnarray}
|L|^2+|T|^2 & = & 1 \nonumber \\
L \, \bar{T} + T \, \bar{L} & = & 0
\end{eqnarray}
In terms of the creation operators ${a_k^\dagger} , \; {d_k^\dagger} , $ and ${s_k^\dagger}$,
Eqs.\,(1) can be rewritten in the form
\begin{eqnarray}
a_k^\dagger & = & \bar{T} \, a_{k-1}^\dagger + \bar{L} \, d_k^\dagger \;,
\nonumber \\
s_k^\dagger & = & \bar{L} \, a_{k-1}^\dagger + \bar{T} \, d_k^\dagger \; , \; \; \; k=1,\; 2, \; \ldots , \; M \; ,
\end{eqnarray}
The input and output modes are connected by a unitary transformation that can be
written in the form of a linear map
\begin{equation}
  \begin{bmatrix}
   a_0^\dagger \\
   d_1^\dagger \\
   d_2^\dagger \\
   d_3^\dagger \\
   \vdots \\
   d_M^\dagger
  \end{bmatrix}
  = U
    \begin{bmatrix}
   a_M^\dagger \\
   s_1^\dagger \\
   s_2^\dagger \\
   s_3^\dagger \\
   \vdots \\
   s_M^\dagger
  \end{bmatrix} \; \; .
\end{equation}
Using Eqs.\,(3), the $(M+1)\times (M+1)$ unitary matrix $U$ can be written
explicitly in terms of the reflection and transmission coefficients of the
beam splitters:
\begin{equation}
  U= 
  \begin{bmatrix}
   T^M & L & LT & LT^2 & \cdots &  LT^{M-1} \\
   LT^{M-1} & T & L^2 & L^2 T & \cdots & L^2 T^{M-2} \\
   LT^{M-2} & 0 & T & L^2 & \cdots & L^2 T^{M-3} \\
   LT^{M-3} & 0 & 0 & T & \cdots & L^2 T^{M-4} \\
   \vdots & \vdots & \vdots & \vdots & \ddots & \vdots \\
   L & 0 & 0 & 0 & \cdots & T
  \end{bmatrix}\; \; .
\end{equation}
It can be checked by straightforward calculation that
the unitarity relations Eqs.\,(2) imply
\begin{equation}
U^\dagger U = \mathbb{I}
\end{equation}
Since by Eqs.\,(4)---(5)
\begin{equation}
a_0^\dagger = T^M a_M^\dagger + L s_1^\dagger + L T s_2^\dagger + L T^2 s_3^\dagger + \cdots +
LT^{M-1} s_M^\dagger \; ,
\end{equation}
an incoming number Fock-state purely in the input mode
\begin{equation}
| \psi_{\rm in} \rangle = |N\rangle = \frac{1}{\sqrt{N!}} (a_0^\dagger)^N|0\rangle
\end{equation}
is transformed to the outgoing state
\begin{equation}
|\psi_{\rm out}\rangle =
\frac{1}{\sqrt{N!}} \left(
T^M a_M^\dagger + L s_1^\dagger + L T s_2^\dagger + L T^2 s_3^\dagger + \cdots +
LT^{M-1} s_M^\dagger \right)^N
|0\rangle \; .
\end{equation}
As mentioned before, Eq.\,(8) assumes that the auxiliary
input modes $\{ d_1 , d_2 , \cdots , d_M \}$ are vacuum ports.
More generally, our model can accommodate noise in the form
of stray photons leaking {\it into} the propagation
channel by simply replacing the input state Eq.\,(8) with a state of the form
\[
| \psi_{\rm in} \rangle = \frac{1}{\sqrt{N!}} (a_0^\dagger)^N
\sum_{q_1 , q_2 , \cdots , q_M} c_{q_1 , q_2 , \cdots , q_M} (d_1^\dagger)^{q_1}
(d_2^\dagger)^{q_2} \cdots (d_M^\dagger)^{q_M}
\; \; |0\rangle
\tag{8'}
\]
To handle incoherent (such as thermal) input noise, one would have to enlarge
the Hilbert space of the auxiliary input modes $\{ d_1 , d_2 , \cdots , d_M \}$
to include their own environment modes, and
trace over these secondary environment modes at the end of
the calculation~\cite{future}.

Going back to the input number state
in the form Eq.\,(8), we can calculate the final output state (density matrix) in the out-mode
(whose creation operator is $a_M^\dagger$) by tracing over the
loss (scattering) modes $s_1 , \, s_2 , \, \ldots , s_M $:
\begin{equation}
\rho_{\rm out} = {\rm{Tr}}_{\{s_1 , \, s_2 , \, \ldots , s_M \}} 
|\psi_{\rm out}\rangle \langle \psi_{\rm out}| \; .
\end{equation}
Substituting Eq.\,(9) for $|\psi_{\rm out}\rangle$, this calculation gives
\begin{equation}
\rho_{\rm out} = \sum_{\substack{ n_0 , \ldots , n_M =0 \\
                                  \sum_{\alpha =0}^{M} n_\alpha = N }}^N
                                  \frac{N!}{n_0 ! n_1 ! \ldots n_M !} \; 
|T|^{2 n_0 M} \; |T|^{2 \sum_{i=1}^M (i-1) n_i }\;
|L|^{2 \sum_{i=1}^M n_i } \; |n_0 \rangle_{s_0} \langle n_0 |_{s_0} \; ,
\end{equation}
where, for ease of combinatorial manipulation, we renamed the output
$M$-mode with index 0: $s_0 \equiv a_M $ (equivalently, $s_0^\dagger = a_M^\dagger$),
and $|k \rangle_{s_0} \equiv |k \rangle_{s_M} $.
It is convenient to combinatorially manipulate Eq.\,(11)
and rewrite it in the following form
\begin{equation}
\rho_{\rm out} = \sum_{n_0 = 0}^N
\frac{N!}{n_0 ! (N-n_0 )!} |n_0 \rangle_{s_0} \langle n_0 |_{s_0}
\sum_{\substack{ n_1 , \ldots , n_M =0 \\
                                  \sum_{j = 1}^{M}
                                   n_j = N-n_0 }}^N
                                  \frac{(N-n_0)!}{n_1 ! n_2 ! \ldots n_M !} \; 
|T|^{2 n_0 M} \; |T|^{2 \sum_{i=1}^M (i-1) n_i }\;
|L|^{2 \sum_{i=1}^M n_i } \; \; .
\end{equation}

\begin{figure}[htp]
%\vspace{0.5in}
\hspace{3.0in}
\centerline{
\input epsf
\setlength{\epsfxsize}{4.2in}%4.200
\setlength{\epsfysize}{3.0in}%3.245
\epsffile{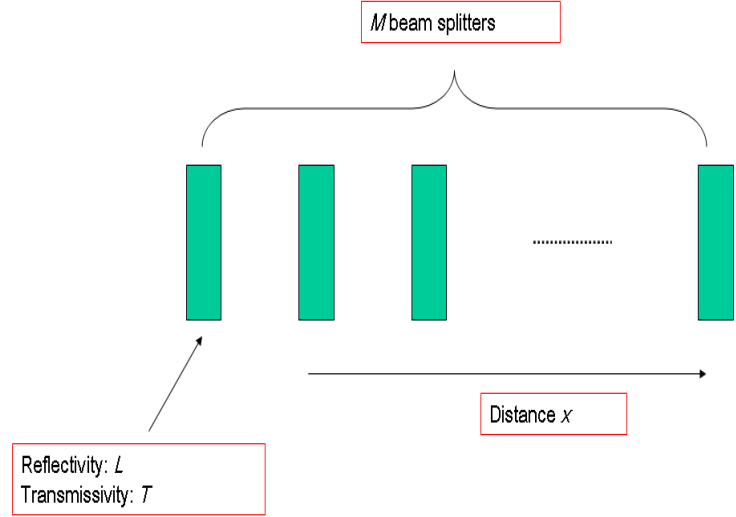}
}
%\vspace{0.2in}
\caption[figure]{\label{fig:figure2}
The ``Beer" limiting process where discrete beam splitters merge
into a continuous propagation medium.}
\vspace{0.10in}
\end{figure}

The continuum limit is defined by taking the number of beam splitters
$M$ to infinity, while keeping the ``power lost" per unit length of propagation
constant. Since each beam-splitter's contribution to power loss is given
by $|L|^2$, mathematically this amounts to the limiting process (Fig.\,2)
\begin{equation}
\mbox{ {\color{red} Limit process for amplitude evolution:\;\;\;}}
\begin{cases}
M \longrightarrow \infty \\
L \longrightarrow 0 \\
|L|^2 \frac{M}{x} \longrightarrow {\rm constant} \equiv \mu \\
|T|^{2M} = (1-|L|^2)^M \longrightarrow
\left(1 - \frac{\mu x}{M} \right)^M \longrightarrow e^{-\mu x}
\end{cases}
\end{equation}
Similarly, to preserve unitarity, a corresponding limit process must govern
the evolution of the ``phase" $\eta$ of the complex transmission amplitude $T$:
\begin{equation}
\mbox{ {\color{red} Limit process for phase evolution:\;\;\;}}
\begin{cases}
M \longrightarrow \infty \\
T \equiv |T| e^{i \phi} \\
\phi \longrightarrow 0 \\
\phi \frac{M}{x} \longrightarrow {\rm constant} \equiv \eta \\
T^{M} \longrightarrow |T|^M e^{i \eta x}
\end{cases}
\end{equation}
Substituting Eqs.\,(13) and (14) into Eq.\,(12) yields our sought-for result
for the $x$-dependent output density matrix resulting from an input Fock
number state with $N$ photons:
\begin{equation}
\rho_{\rm out} (x) = \sum_{n=0}^N \binom{N}{n}
e^{-n \mu x} (1- e^{ -\mu x})^{N-n} \;
|n \rangle \langle n | \; .
\end{equation}
The interpretation of Eq.\,(15) in terms of photon loss as a function of propagation
distance is straightforward (see also~\cite{durkinth} for an alternate derivation
with a fixed amount of loss).

~

\begin{figure}[htp]
%\vspace{0.5in}
\hspace{3.0in}
\centerline{
\input epsf
\setlength{\epsfxsize}{4.0in}%4.200
\setlength{\epsfysize}{3.245in}%3.245
\epsffile{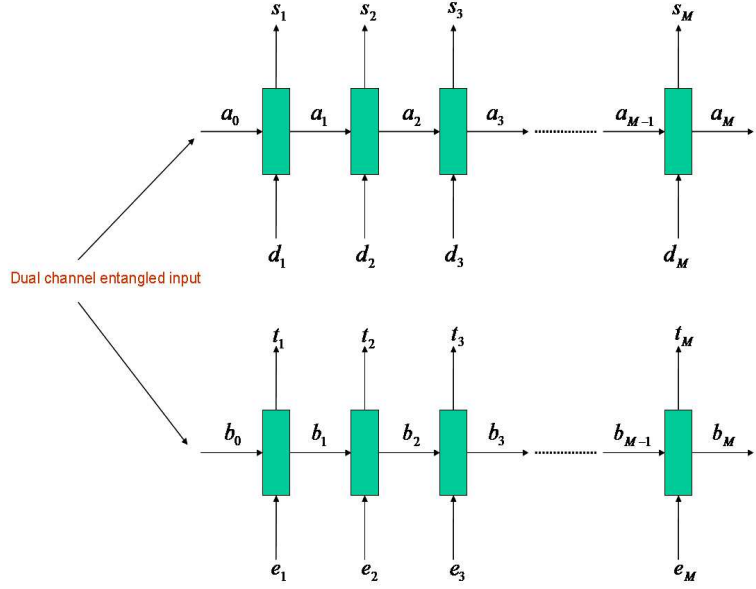}
}
%\vspace{0.3in}
\caption[figure]{\label{fig:figure3}
Modeling propagation loss for dual-channel entangled
photon states with a series of beam splitters.}
\vspace{-0.0in}
\end{figure}

Let's now turn to the analysis of dual-channel lossy propagation.
We proceed similarly to the single-mode case, and adopt the notation
for labeling the input, output, and scattering modes as illustrated by Fig.\,3.
Everything in the analysis above between Eqs.\,(1) and (7) proceeds independently for the
two propagation channels $a$ and $b$, with corresponding series of beam splitters
having in general distinct reflection and transmission characteristics.
The dual-channel version of Eq.\,(7) is
\begin{eqnarray}
a_0^\dagger & = & T_a^M a_M^\dagger + L_a s_1^\dagger + L_a T_a s_2^\dagger + L_a T_a^2 s_3^\dagger + \cdots +
L_a T_a^{M-1} s_M^\dagger \nonumber \\
b_0^\dagger & = & T_b^M b_M^\dagger + L_b t_1^\dagger + L_b T_b t_2^\dagger + L_b T_b^2 t_3^\dagger + \cdots +
L_b T_b^{M-1} t_M^\dagger \; .
\end{eqnarray}
As a first, important example, consider an input N00N state which has the
form
\begin{equation}
|\psi_{\rm in} \rangle = \frac{1}{\sqrt{2}}
\left( |N \rangle_a |0 \rangle_b + |0 \rangle_a |N \rangle_b \right)
=\frac{1}{\sqrt{2 N!}} \left(
(a_0^\dagger )^N + (b_0^\dagger)^N \right) |0 \rangle_a | 0 \rangle_b \; .
\end{equation}
According to Eq.\,(16), the output state is given by
\begin{eqnarray}
|\psi_{\rm out}\rangle & = & \frac{1}{\sqrt{2 N!}} \left[ \left(
T_a^M a_M^\dagger + L_a s_1^\dagger + L_a T_a s_2^\dagger + L_a T_a^2 s_3^\dagger + \cdots +
L_a T_a^{M-1} s_M^\dagger \right)^N \right. \nonumber \\
& + & \left. \left(
T_b^M b_M^\dagger + L_b t_1^\dagger + L_b T_b t_2^\dagger + L_b T_b^2 t_3^\dagger + \cdots +
L_b T_b^{M-1} t_M^\dagger \right)^N \right] |0 \rangle_a |0 \rangle_b \; \; .
\end{eqnarray}
As before, the output density matrix is obtained by tracing
over the scattering modes:
\begin{equation}
\rho_{\rm out} = {\rm{Tr}}_{\{s_1 , \, \ldots , s_M , t_1 ,  \ldots , t_M\}} 
|\psi_{\rm out}\rangle \langle \psi_{\rm out}| \; .
\end{equation}
Substituting Eq.\,(18) in Eq.\,(19) and carrying out some straightforward algebra,
we reach the dual-channel analogue of Eq.\,(12), which we will suppress here to save space.
Applying the Beer limit process ($M \rightarrow \infty$)
defined by Eqs.\,(13) and (14) to this expression
(whereby each channel $a$, $b$ is associated with its own version of the extinction and
phase-rotation coefficients $\mu_a$, $\mu_b$, $\eta_a$, and $\eta_b$) yields
the following continuum limit for the output density matrix of a
N00N input state as a function of propagation distance $x$:
\begin{eqnarray}
\rho_{\rm out} (x)
& = & \frac{\; 1 \;}{\; 2 \;} \, \BigParLeft \! \! \! \sum_{n=0}^N \binom{N}{n} \left[ \,
e^{-n \mu_a x} (1- e^{ -\mu_a x})^{N-n} \; |n \rangle_a |0 \rangle_b
\langle n|_a \langle 0 |_b
+  e^{-n \mu_b x} (1- e^{ -\mu_b x})^{N-n} \; |0 \rangle_a |n \rangle_b
\langle 0|_a \langle n |_b \; \right] \nonumber \\
& + & 
e^{-\frac{N}{2} (\mu_a + \mu_b ) x} \, \left(
e^{i (\eta_a - \eta_b )x } |N \rangle_a |0 \rangle_b \langle 0 |_a \langle N|_b
+ e^{-i (\eta_a - \eta_b )x } |0 \rangle_a |N \rangle_b \langle N |_a \langle 0|_b
\right) \,
\BigParRight \; \; .
\end{eqnarray}
Note that the coherence term is the entire expression on the second line
of Eq.\,(20). This term is exponentially suppressed with an extinction coefficient
given by $N \mu$, where $\mu \equiv (\mu_a + \mu_b) /2$ is the average extinction
coefficient of the two propagation channels. Hence the well-understood ``super-exponential"
loss of coherence (hence loss of entanglement) in the propagation of N00N states
is once again verified~\cite{N00N-loss}.

After working through the N00N state example above, it is straightforward but
rather cumbersome to extend our calculation to a completely general two-mode input state,
the only restriction being that the maximum photon number $N$ in each mode $a$ and $b$
is finite. This most general input state can be written in the form
\begin{equation}
| \psi_{\rm in} \rangle = \sum_{l,\, m=0}^{N}
\frac{\alpha_{lm}}{\sqrt{l! m!}} ({a_0}^{\dagger})^l
({b_0}^{\dagger})^m |0 \rangle_a \, |0 \rangle_b \; , \; \; \; \; \; \;
\sum_{l, \, m=0}^N |\alpha_{lm}|^2 = 1 \;.
\end{equation}
Straightforward calculation in the spirit of the analysis presented thus far
gives the result
\begin{eqnarray}
\rho_{\rm out}(x_a , x_b ) =
\sum_{l,\, m, \, l' , \, m' =0}^{N}
\alpha_{lm} \overline{\alpha_{l'm'}}
e^{i(l-l') \eta_a x_a} e^{i (m-m') \eta_b x_b}
e^{-\tfrac{1}{2} (l'-l) \mu_a x_a} e^{-\tfrac{1}{2} (m'-m) \mu_b x_b} \times \nonumber \\
\times
\sum_{p=0}^{l} \sum_{q=0}^{m}
\frac{1}{(l-p)! (m-q)!} \left( \frac{l! \; l'! \; m! \; m'!}
{p! q! (p+l'-l)! (q+m'-m)!} \right)^{\tfrac{1}{2}} \times \nonumber \\
\times \;
e^{-p \mu_a x_a} (1-e^{-\mu_a x_a})^{l-p}\,
e^{-q \mu_b x_b} (1-e^{-\mu_b x_b})^{m-q}\,
|p\rangle_a |q\rangle_b \langle p+l'-l |_a \langle q+ m'-m |_b  \; \; ,
\end{eqnarray}
where we also implemented a trivial generalization by allowing the output ``point"
to have different $x$ coordinates $x_a$ and $x_b$ along the two distinct propagation
paths. Here and in what follows we adopt the combinatorial convention that the factorial
of a negative integer is $+\infty$; thus the contributions to the above sum from, e.g.,
terms with $l < p$ or $p+l' < l$ vanish by virtue of the factorial terms in the denominator.
A somewhat lengthy binomial-chase through the sums in Eq.\,(22) allows us
to put it into the slightly more manageable alternative form:
\begin{eqnarray}
\rho_{\rm out}(x_a , x_b )
& = & \sum_{p,\, q,\, p',\, q'=0}^{N}
|p \rangle_a |q \rangle_b \langle p' |_a \langle q' |_b \times \nonumber \\
& \times & e^{-\tfrac{1}{2} (p+p') \mu_a x_a} e^{-\tfrac{1}{2} (q+q' ) \mu_b x_b}
e^{i(p-p' )\eta_a x_a} e^{i (q-q') \eta_b x_b} \times \nonumber \\
& \times & \sum_{l=p}^{N} \sum_{m=q}^N \frac{\alpha_{lm} \overline{\alpha_{l+p'-p \; m+q'-q}}}
{(l-p)! (m-q)!} \left( \frac{l! (l+p'-p)! \, m! (m+q'-q)! }
{p! \, q! \, p'! \, q'!} \right)^{\tfrac{1}{2}} \times \nonumber \\
& \times&  (1-e^{-\mu_a x_a})^{l-p} \, (1-e^{- \mu_b x_b})^{m-q} \; \; \; \;.
\end{eqnarray}

Before we begin analyzing the consequences of our main result Eq.\,(23)
quantitatively, we mention one last generalization which is quite
straightforward. It is easy to incorporate $x$-dependent (variable) extinction
and phase rotation coefficients $\mu$ and $\eta$ into the above development
merely via the substitutions in the limit process Eqs.\,(13)--(14)
\begin{eqnarray}
e^{-\mu x} & \longrightarrow & e^{- \int_0^x \mu (\zeta) \, d \zeta} \nonumber \\
e^{i \eta x} & \longrightarrow & e^{i \int_0^x \eta (\zeta) \, d \zeta} \; \; \; .
\end{eqnarray}
With this generalization, the result Eq.\,(23) becomes
\begin{eqnarray}
\rho_{\rm out}(x_a , x_b )
& = & \sum_{p,\, q,\, p',\, q'=0}^{N}
|p \rangle_a |q \rangle_b \langle p' |_a \langle q' |_b \times \nonumber \\
& \times & e^{-\tfrac{1}{2} (p+p') \int_0^{x_a} \mu_a }
\; e^{-\tfrac{1}{2} (q+q' ) \int_0^{x_b}\mu_b}\;
e^{i(p-p' )\int_0^{x_a}\eta_a} \; e^{i (q-q') \int_0^{x_b}\eta_b } \; \times \nonumber \\
& \times & \sum_{l=p}^{N} \sum_{m=q}^N \frac{\alpha_{lm} \overline{\alpha_{l+p'-p \; m+q'-q}}}
{(l-p)! (m-q)!} \left( \frac{l! (l+p'-p)! \, m! (m+q'-q)! }
{p! \, q! \, p'! \, q'!} \right)^{\tfrac{1}{2}} \times \nonumber \\
& \times&  (1-e^{-\int_0^{x_a}\mu_a })^{l-p} \; (1-e^{- \int_0^{x_b}\mu_b })^{m-q} \; \; \; \;.
\end{eqnarray}

We now calculate the decay of coherence and entanglement numerically
using the above formalism for N00N and generic
(random) entangled states. The main results are illustrated by the plots
in Figs.\,4--7.

\begin{figure}[htp]
%\vspace{0.5in}
\hspace{3.0in}
\centerline{
\input epsf
\setlength{\epsfxsize}{4.0in}%4.200
\setlength{\epsfysize}{3.245in}%3.245
\epsffile{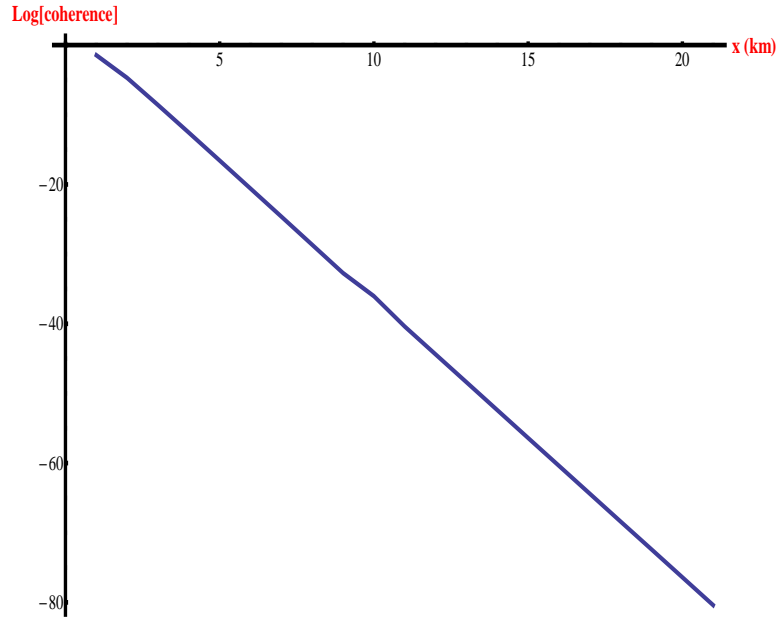}
}
%\vspace{0.0in}
\caption[figure]{\label{fig:figure4}
Decay behavior of coherence power (the sum of absolute
squares of the off-diagonal elements) for a $N=10$ N00N state.}
\vspace{-0.0in}
\end{figure}

All of our plots are generated assuming constant
extinction and rotation coefficients $\mu = \mu_a = \mu_b = 0.2 {\rm km}^{-1}$
and $\eta = \eta_a = \eta_b = 1 {\rm km}^{-1}$.
The amount of coherence is calculated by computing the sum
of absolute squares of the off-diagonal
elements of the density matrix (the off-diagonal matrix norm squared); hence
it can be interpreted as the coherence ``power" that survives in the density
matrix as it propagates through the lossy medium. Also note that in all
three plots Figs.\,4--7 the $x$ axis is the propagation distance,
and the $y$-axis depicts the logarithm of the
dependent variable.

Fig.\,4 makes clear the ``super-exponential" decay of coherence
for a N00N state as a function of propagation distance.
For the $N=10$ N00N state, coherence power decays
proportionally to $\exp (-2 N \mu x )$ as expected from Eq.\,(20).

%
%\begin{figure}[htp]
%\vspace{-0.5in}
%\hspace{3.0in}
%\centerline{
%\input epsf
%\setlength{\epsfxsize}{4.0in}%4.200
%\setlength{\epsfysize}{3.245in}%3.245
%\epsffile{fig5.eps}
%}
%\vspace{-0.3in}
%\caption[figure]{\label{fig:figure5}
%Decay of coherence power for a random entangled state (Eq.\,(21))
%with $N=10$. Note the slight plateau in decay at medium propagation distances.}
%\vspace{0.0in}
%\end{figure}
%

%
\begin{figure}[htp]
%\vspace{0.5in}
\hspace{3.0in}
\centerline{
\input epsf
\setlength{\epsfxsize}{4.0in}%4.200
\setlength{\epsfysize}{3.245in}%3.245
\epsffile{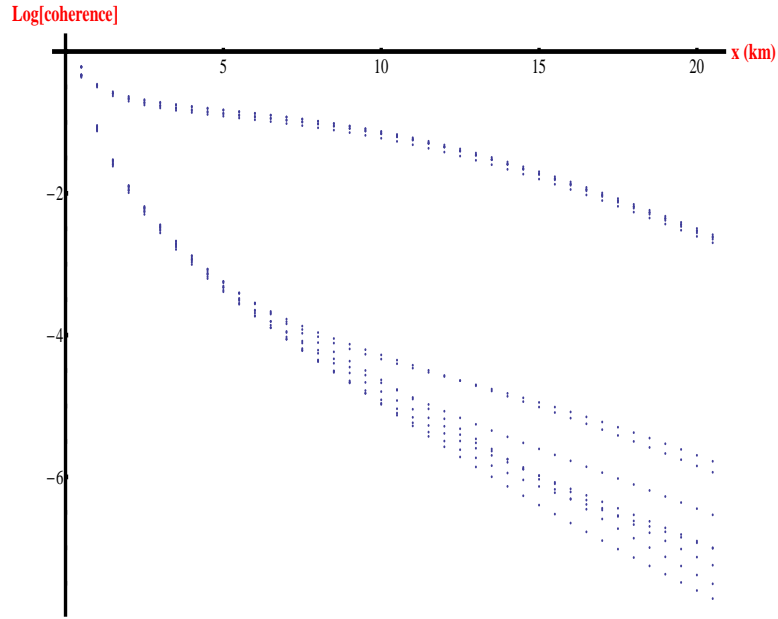}
}
%\vspace{0.0in}
\caption[figure]{\label{fig:figure5}
Decay of coherence power for two ensembles of random entangled states (Eq.\,(21))
with $N=10$. Note the slight plateau in decay at medium propagation distances.
The plateau reverts back to exponential
decay at large propagation distances; however, the overall decay remains no worse
than that of classical light. The upper ensemble is drawn from a joint uniform
distribution for the real and imaginary parts of $\alpha_{ij}$, which
is then normalized. The lower ensemble is drawn from a true uniform distribution
on the Bloch sphere in the complex space ${\BC}^{N+1}$, where here $N=10$.}
\vspace{-0.0in}
\end{figure}

In contrast to Fig.\,4, Fig.\,5 suggests an unexpected robustness of coherence
in generic entangled states
against propagation loss. Here we 
plot the coherence power as a function
of propagation distance for two ensembles
of entangled states, drawn randomly from two different
probability distributions on the Hilbert space
of all states of the form Eq.\,(21).
One ensemble is drawn from a joint uniform
distribution for the real and imaginary parts of $\alpha_{ij}$, which
is then normalized. The other ensemble is drawn from a true uniform distribution
on the Bloch sphere in the complex space ${\BC}^{N+1}$.
To generate the latter ensemble, we utilize the result
that $(z_1 , \cdots z_n )/\sqrt{|z_1 |^2 + \cdots + |z_n |^2}$
is a random vector on the unit sphere $S^{2n-1}$ in ${\BC}^{n}$
with respect to the canonical
volume form on $S^{2n-1}$ if ${\rm{Re}}(z_i )$, ${\rm{Im}} (z_j )$
are i.i.d.\ Gaussian random variables with zero mean~\cite{mars,mull}.
We see from Fig.\,5 that:
\begin{itemize}
\item For a random entangled input state, coherence decays no
faster than the classical loss rate.\\

\item There is a mid-range plateau in loss, where decay is
sub-classical, after which classical decay-rate resumes.
This feature is present for all generic entangled inputs.\\

\item How much of the surviving generic coherence is useful for
super-classical resolution will depend more on the survival rate
of entanglement than on that of coherence.
\end{itemize}

\begin{figure}[htp]
%\vspace{-0.5in}
\hspace{3.0in}
\centerline{
\input epsf
\setlength{\epsfxsize}{4.0in}%4.200
\setlength{\epsfysize}{3.245in}%3.245
\epsffile{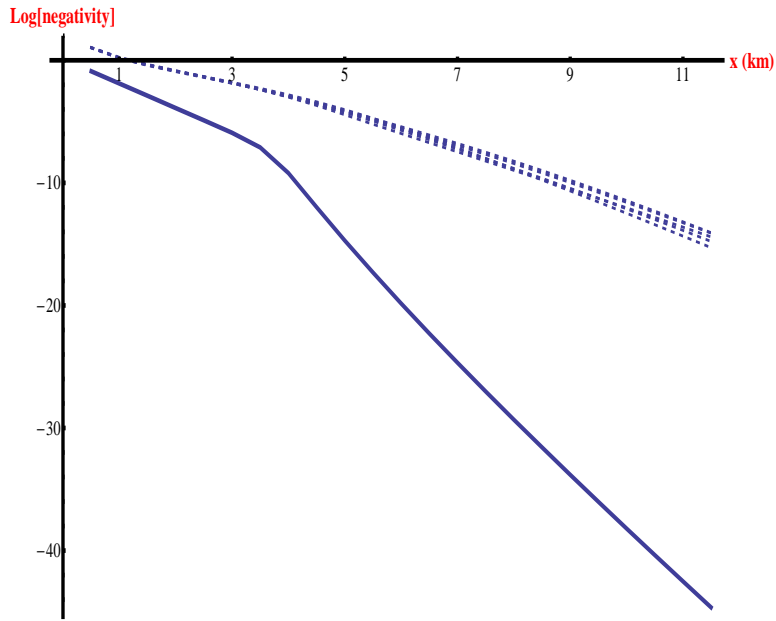}
}
%\vspace{0.0in}
\caption[figure]{\label{fig:figure6}
Behavior of negativity for the $N=10$
N00N state and a random ensemble of entangled states ($N=10$).
The ensemble is drawn from a true uniform distribution
on the Bloch sphere in the complex space ${\BC}^{N+1}$ as in Fig.\,5.}
\vspace{-0.0in}
\end{figure}

To investigate the decay of entanglement in the density matrix that results from
the propagation of a generic input state of the form Eq.\,(21),
we use the concept of ``negativity"~\cite{negativity,nielsen}. Consider
a mixed state of a general bi-partite system with Hilbert space
$\SH_A \otimes \SH_B$ given by
\begin{equation}
\rho = \sum_{i,j,k,l} \rho_{ij,kl}|i_A j_B\rangle \otimes \langle k_A l_B | \; ,
\end{equation}
where $\{ |i_A j_B\rangle \}$ is a separable orthonormal basis for $\SH_A \otimes \SH_B$.
The ``partial transpose" of $\rho$ with respect to the subsystem $A$
is defined as the operator
\begin{equation}
\rho^{t_A} \equiv \sum_{i,j,k,l} \rho_{kj,il}|i_A j_B\rangle \otimes \langle k_A l_B | \; .
\end{equation}
Although the partial transpose $\rho^{t_A}$ is symmetric and has unit trace,
it is in general not a density matrix
since it is not necessarily a positive operator. In fact, the separability
of $\rho$ is a sufficient (though not necessary)
condition for the positivity of $\rho^{t_A}$. In other words, the negativity
of $\rho^{t_A}$ guarantees entanglement, but it is not necessary that $\rho^{t_A}$
is negative for entanglement to be present. This makes negativity a partial
entanglement measure which is relatively easy to compute. Quantitatively,
negativity is defined to be the (non-negative) quantity
\begin{equation}
\sum_{i: \lambda_i^{t_A} < \; 0} | \lambda_i^{t_A} | \; ,
\end{equation}
where the sum is over all eigenvalues of $\rho^{t_A}$ which are negative.

In Fig.\,6 we illustrate the behavior of negativity as a function
of propagation distance for a N00N state and for a random ensemble
of entangled states drawn randomly from the uniform
probability distribution on the Bloch sphere in the Hilbert space
of all states of the form Eq.\,(21). (See~\cite{durkin} for
a different look at the behavior of entanglement under loss.)

It is apparent that the ``super-exponential"
propagation loss behavior of N00N states is a highly special
property not shared by more general two-mode entangled non-classical
states of light. Whether the latent robust entanglement of generic two-mode states
illustrated in Fig.\,6 can
be used to reach super-classical resolution in metrology applications
depends on what detection schemes can be deployed on the output
density matrix~\cite{newp}, as well as, from a practical point of view, on whether the
generic states can be created reproducibly. These are questions
we will investigate in a forthcoming paper~\cite{future}.

~~~~

\begin{acknowledgments}
I am grateful to Gabriel Durkin for useful correspondence
on the quantitative description of entanglement in mixed states.
The research work reported on in this paper
was carried out with support from
the Defense Advanced Research Project Agency.
The views, opinions, and findings contained in this article are those of the author
and should not be interpreted as representing the official views or policies,
either expressed or implied, of the Defense Advanced Research Projects Agency
or the Department of Defense.
\end{acknowledgments}

\end{document}